\documentclass[10pt,conference,letterpaper]{IEEEtran}
\usepackage[T1]{fontenc}
\usepackage[utf8]{inputenc}
\usepackage[table]{xcolor}
\usepackage{todonotes}
\usepackage{letltxmacro}
\usepackage{graphicx}
\usepackage{xparse}
\usepackage{xspace}
\usepackage{chngcntr}
\usepackage{caption} 
\usepackage{ucs} 
\usepackage{tablefootnote}
\usepackage{times}
\usepackage{tikz}

\ifdefined\isabridged
\usepackage[bookmarks=false]{hyperref}
\else
\usepackage{hyperref}
\fi

\newif\ifabridged
\newif\ifnotabridged
\newif\ifanonymous
\newif\ifnotanonymous

\ifdefined\isabridged
\abridgedtrue
\fi

\ifabridged\notabridgedfalse
\else\notabridgedtrue
\fi

\ifdefined\isanonymous
\anonymoustrue
\fi

\ifanonymous\notanonymousfalse
\else\notanonymoustrue
\fi

\LetLtxMacro{\todonote}{\todo}
\renewcommand{\todo}[2][]
{\todonote[caption={#2}, size=\footnotesize, #1]
{\renewcommand{\baselinestretch}{0.5}\selectfont#2\par}}

\ifanonymous
\DeclareRobustCommand\sectt[1]{{\fontsize{13}{12}#1}}
\newcommand{\openteesectitle}{\protect\sectt{TEE-Vi}\xspace}
\newcommand{\opentee}{\protect{TEE-Vi}\xspace}
\else
\DeclareRobustCommand\sectt[1]{{\fontsize{13}{12}#1}}
\newcommand{\openteesectitle}{\protect\sectt{Open-TEE}\xspace}
\newcommand{\opentee}{\protect{Open-TEE}\xspace}
\fi

\counterwithin*{paragraph}{subsection}

\ifabridged
\fi

\ifanonymous
\title{TEE-Vi - A Virtual Trusted Execution Environment}
\else
\title{Open-TEE - An Open Virtual Trusted Execution Environment}
\fi

\ifanonymous
\else
\author{
  \IEEEauthorblockN{Brian McGillion\IEEEauthorrefmark{1}, Tanel Dettenborn\IEEEauthorrefmark{2}, Thomas Nyman\IEEEauthorrefmark{3}}
  \IEEEauthorblockA{Intel Collaborative Research Institute for Secure
    Computing (ICRI-SC) at Aalto University, Finland.}\\
  \IEEEauthorblockN{N. Asokan\IEEEauthorrefmark{4}}
  \IEEEauthorblockA{Aalto University and University of Helsinki, Finland.}\\
  \IEEEauthorrefmark{1}brian.mcgillion@intel.com,
  \IEEEauthorrefmark{2}tanel.dettenborn@intel.com,
  \IEEEauthorrefmark{3}thomas.nyman@aalto.fi,
  \IEEEauthorrefmark{4}asokan@acm.org.
}

\fi

\begin{document}

\maketitle
\begin{abstract}
Hardware-based Trusted Execution Environments (TEEs) are widely
deployed in mobile devices. Yet their use has been limited primarily
to applications developed by the device vendors. Recent
standardization of TEE interfaces by GlobalPlatform (GP) promises to
partially address this problem by enabling GP-compliant trusted
applications to run on TEEs from different vendors. Nevertheless
ordinary developers wishing to develop trusted applications face
significant challenges. Access to hardware TEE interfaces are difficult
to obtain without support from vendors. Tools and software needed to
develop and debug trusted applications may be expensive or
non-existent.

In this paper, we describe \opentee\ifanonymous\footnotemark\fi, a
virtual, hardware-independent TEE implemented in software. \opentee
conforms to GP
specifications. It allows developers to develop and debug trusted
applications with the same tools they use for developing software in
general. Once a trusted application is fully debugged, it can be
compiled for any actual hardware TEE. Through performance measurements
and a user study we demonstrate that \opentee is efficient and easy to use.
%We also describe our experience
%in getting trusted application developers to try out \opentee.
\ifanonymous
\footnotetext{Name anonymized for submission}
\fi
\ifanonymous
%% Don't mention github in the submission version
\else We have made \opentee freely available as open
source\footnote{\url{https://open-tee.github.io}}.  \fi
\end{abstract}

\ifnotabridged
\begin{tikzpicture}[remember picture,overlay]
  \node[anchor=south,yshift=10pt] at (current page.south) {\fbox{\parbox{\dimexpr\textwidth-\fboxsep-\fboxrule\relax}{
    \footnotesize This is the author's version of the article to appear in 14th
    IEEE International Conference on Trust, Security and Privacy in Computing
    and Communications, TrustCom 2015, Helsinki, Finland, August 20-22, 2015.
  }}};
\end{tikzpicture}%
\fi

% Introduction
\section{Introduction}
\label{sec:intro}

Personal computing devices such as smartphones, tablets and laptops have become pervasive. They are used to store sensitive data and access critical services across a wide range of domains, such as banking, health care and safety, where privacy and security are paramount. On the other hand, traditional operating systems and the services that they provide are becoming so large and complex that the task of securing them is increasingly harder.  Hardware-based trusted execution environments (TEEs) were developed to address this gap. A TEE on a device is isolated from its main operating environment by using hardware security features. It offers a smaller operating environment that provides just enough functionality so that sensitive data and operations can be offloaded to it.  %There is generally no need for large run-times and complex libraries with a myriad of inter-dependencies that are derived from a series of potentially 3rd party vendors, as is the case in a standard OSs such as Microsoft Windows.  With the increased demand for privacy and security among users, for banking, medical and safety critical devices the need for the added security offered by the TEE is becoming paramount.

Hardware-based TEEs have been widely deployed in mobile devices for over a decade~\cite{EkbergKA14}. TI M-Shield~\cite{white:mshield} and ARM TrustZone~\cite{web:TRUSTZONE,web:ARM_TZ} are early examples, followed by newer architectures like the Intel SEP security co-processor~\cite{web:SEP} and Apple's ``Secure Enclave'' co-processor~\cite{web:APPLE_SE}. Business requirements such as the need to enforce digital rights management and subsidy locks, as well as regulatory requirements like cloning- and theft protection have been the driving forces behind such large scale deployment~\cite{EkbergKA14}. Such requirements continue to appear: e.g., fingerprint scanners with hardware protection, hardware-backed keystores, and the recent ``kill switch''~\cite{web:SB962} bill in California mandating that a mobile device must be capable of being rendered inoperable if it is stolen.

Although the early hardware security modules (HSMs) like the IBM cryptocards\footnote{\url{http://www.ibm.com/security/cryptocards/}} were programmable~\cite{Dyer+01}, the vast majority of HSMs used with personal computers and servers today are typically application-specific modules or fixed function co-processors like the Trusted Platform Modules (TPMs)~\cite{TPM}. In contrast, TEEs in mobile devices are programmable. However, despite widespread deployment of hardware-based TEEs in mobile devices, application developers have lacked the interfaces to use TEE functionality to protect their applications and services. Nor have they been researched extensively in the academic community.  Recent efforts by GlobalPlatform~\cite{web:GP} to specify standard interfaces for TEE functionality in mobile devices~\cite{web:GP_TEE} will partially address this problem. However, there are a number of factors that stand in the way of widespread use of hardware-based TEEs in application development and research. Chief among them is the difficulty of developing applications for TEEs. Software development kits for TEE application development are often proprietary or expensive. Debugging low-level TEE applications either requires expensive hardware debugging tools, or leaves the developer with only primitive debugging techniques like ``print tracing'' (e.g., using printf statements in C to keep track of how values of variables change during program execution).  

In this paper, we argue that a virtual standards-compliant TEE implemented entirely in software will allow developers to build TEE applications using tools and development environments that they are \textit{already} familiar with. It will also allow applications to be tested and refined even when developers do not have access to devices where hardware TEE functionality has been made accessible to them. Such a facility will greatly ease TEE application development and can trigger new ways of using TEEs. We make the following contributions:
\begin{itemize}
\item We design and implement such a \textbf{virtual TEE called \opentee} which conforms to GlobalPlatform Specifications. We identify requirements that would make \opentee acceptable to developers and make specific design choices informed by these requirements (Section~\ref{sec:opentee}). \opentee is publicly available on GitHub.\ifanonymous\footnote{URL removed for blind reviewing}\else\footnote{\url{http://open-tee.github.io/}}\fi
\item We show that \opentee is \textbf{efficient, hardware-independent} and allows a developer to carry out much of the development life cycle of standard-compliant TEE applications using popular application development environments they already use. We demonstrate that \opentee \textbf{significantly improves the ease-of-use} of TEE application development by conducting a small-scale, yet rigorous, user study with experienced professional TEE developers (Section~\ref{sec:eval}). 
\end{itemize}
Given the demonstrable usability benefits, we recommend that organizations who develop applications for TEEs should consider incorporating \opentee into their development process. We also hope that this paper will enable more researchers to discover the power of TEEs and use \opentee to develop and experiment with new TEE applications.

Note that \opentee is \textit{not intended to emulate} a particular hardware TEE. The goal of \opentee is that a Trusted Application developed successfully with \opentee is guaranteed to compile and run on any target GlobalPlatform-compliant TEE.
\section{Background}
\label{sec:background}

\subsection{Structure  of a TEE}
\begin{figure}[tb]
\centering
\includegraphics[width=3.1in]{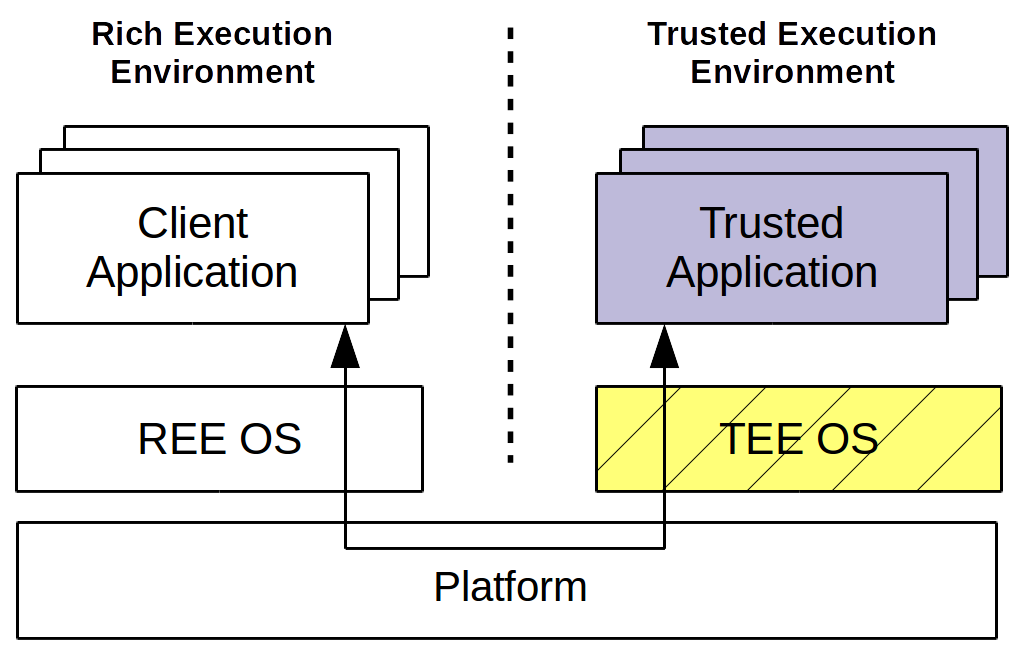}
\caption{A TEE in a Computing Device}
\label{fig:teearch}
\end{figure}

A TEE is a secure, integrity-protected processing environment, consisting of processing, memory and storage capabilities. Figure~\ref{fig:teearch} shows how a device can be visualized as a series of distinct environments with their own set of features and services. Using the terminology introduced by GlobalPlatform~\cite{web:GP_TEE_SA} we describe below the concepts illustrated in Figure~\ref{fig:teearch}:

\noindent\textbf{Rich Execution Environment (REE)}: The word ``rich'' here refers to an operating environment that is feature rich such as one would expect from a modern platforms such as Android, iOS, Windows, Linux or OS X.  %In some literature this environment may be referred to as the ``Normal World'' in reference to the fact that it is where the majority of applications are being developed for and deployed to.
%%
%% NA Note: Normal world/secure world is TZ-specific terminology. Don't introduce it here.

\noindent\textbf{Trusted Execution Environment (TEE)}: The TEE is a combination of features, both software and hardware, that isolate the execution of tasks from the REE.  These environments have a limited set of features and services as they are intended to only address the security critical subset of an application's functionality such as offloading some cryptographic operations or key management.

\noindent\textbf{Trusted Application (TA)}: An application encapsulating the security-critical functionality to be run within the TEE. This may be a service style application that provides a general feature, such as a generic cryptographic keystore, or it could be designed to offload a very specific part of an application that is running in the REE, such as a portion of the client state machine in a security protocol like TLS.

\noindent\textbf{Client Application (CA)}: CAs are ordinary applications (e.g., browser or e-mail client) running in the REE. CAs are responsible for providing the majority of an application's functionality but can invoke TAs to offload sensitive operations.

As an example, consider a common use case for TEEs: the offloading of Digital Rights Management (DRM) protected content. The CA would be responsible for the majority of the tasks associated with viewing the content i.e. opening the media file, providing a region in the display into which it can be rendered (the window) and providing a mechanism to start, stop, rewind the media.  A TA would be used to decrypt the protected media stream and make the decrypted content available directly to the graphics hardware that is responsible for rendering and displaying the stream. 

\ifabridged
\else
\subsection{TEE architectural options}
\label{subsec:arch_options}
\begin{figure*}
\includegraphics[width=\textwidth]{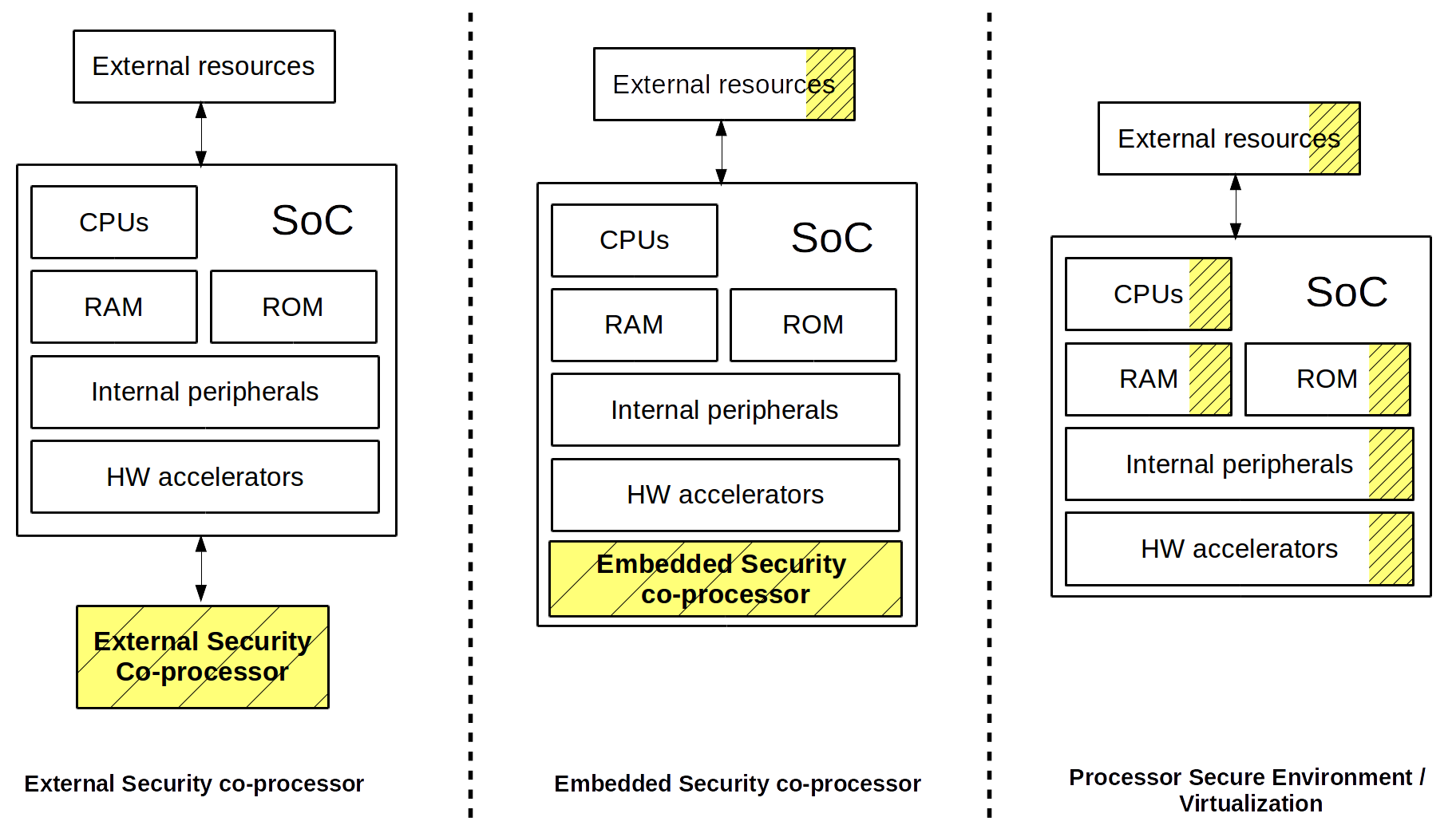}
\caption{Three potential architectural options for realizing a TEE architecture (adapted from\protect~\cite{web:GP_TEE_SA})}
\label{fig:teearch_options}
\end{figure*}

A TEE can be realized in different ways, but the overall concept stays the same.
Figure~\ref{fig:teearch_options} shows a number of ways in which these TEEs can be realized:

\noindent\textbf{Co-Processor}: A separate core, generally with its own peripherals, is used to offload the security critical tasks from the main operating environment.  The benefits of such a configuration are that the operation can generally be completely isolated and it can run simultaneously with the main core.  The drawback is that there is an overhead associated with transferring the data to and from the core. Also, the co-processor is generally less powerful than the main core.  The co-processor design can be further separated into two alternatives:

\noindent\textit{External Security co-processor}: is a discrete hardware module outside the physical chip (commonly referred to as ``System on Chip'' or SoC) containing the main core, and is thus completely isolated from it, not sharing any resources with it.

\noindent\textit{Embedded Security co-processor}: is embedded into the main SoC and thus has the capability to share some of the resources of the main system. It is still isolated from the main processor.

\noindent\textbf{Processor Secure Environment}: Many popular mobile TEE architectures follow a configuration where a single core supports multiple virtual cores that are mutually exclusive of one another i.e. when one is running the other is suspended.  Generally there is some form of trigger to allow the core to switch from one state to the other. This configuration is sometimes referred to as the ``processor secure environment''~\cite{EkbergDiss}.

\noindent\textit{ARM TrustZone} is an example of this configuration. In TrustZone, the processor core can be in one of two ``worlds'': a ``secure world'' (for the TEE) and a ``normal world'' (for the REE). A special instruction called Secure Monitor Call (SMC) can be executed to trigger the processor running in normal world to enter ``monitor mode'' that marshals the transition to secure world~\cite{web:TRUSTZONE}.  The advantage of this configuration is that there is no need to offload the data to and from the secure world. However, there is a cost associated with having to store and restore the device state on entry and exit from a given mode.  On single core devices there is also an added security benefit from having only one world running at a given time in that it ensures that the normal world OS cannot 
%perform a nefarious operation targeted at the secure world during the operation of the secure world i.e. 
interfere with the secure world directly or indirectly (e.g., software side-channel attacks). However, this also has the disadvantage that when one world is active the other world must be completely halted, thus complicating interrupt handling.% and potentially causing a transition back to handle the interrupt before the task is complete.
%%
%% NA Note: Let's not get into too much detail here. AFAIK, TZ operation cannot be pre-empted by interrupts. This is why TZ OSs use a watchdog timer to hand control back to REE.

\noindent\textit{Intel Software Guard Extensions (SGX)}~\cite{SGX13,web:SGX} is another example of such a variant, the core does not perform a full transition to and from a secure world.  Instead parts of a standard application, both code and data, are protected by mechanisms in the core. Parts of the application, called an Enclave, are encrypted by a key that is only accessible to the CPU.  When an ``enter enclave (EENTER)''~\cite{web:SGX2} instruction is received the code and data are decrypted and operated upon in the core. They never leave the CPU package unencrypted, thus protecting them against external access.  The benefits are that there is no need to transfer data back and forth between cores or to setup complicated transitions to and from a secure world, and there is no additional need for a separate operating environment as is required in other styles of TEE configuration.  %Most security use cases that result from normal world applications running can be supported in this fashion by allowing 
%TEE usage scenarios can be realized in this fashion
%small select parts of the application to be secured, thus enabling us to move back from the paradigm of ``Splitting Trust''\cite{conf:balfanz1999hand} 
%which towards a model that developers are more familiar with,  where an application can be self contained.  
%The drawbacks currently with this approach are that the application needs to be customized to take advantage of this feature through the use of a supplied SDK.
%%
%% NA Note: again, I wouldn't call this an advantage of SGX. Even SGX developers have to know how to annotate their code to identify enclaves. The new research by MSR (and what we discussed with Elena) about allowing unmodified apps to benefit from TEEs would achieve this. But SGX and all other TEEs today don't in their current form.

\noindent\textit{Virtualization} based on hardware features such as AMD-V and Intel VT-(x,d), have existed for many years and are used extensively to provide separation of resources between different operating environments especially in high density server configurations.  They rely on processor support to allow virtualization of instructions and access to resources e.g. through the use of an IOMMU\footnote{Input/Output Memory Management Unit}, access to and from peripheral devices can be restricted.  Though in and of themselves they are not designed solely to provide a TEE, there is recent research~\cite{jour:cheng2013appshield,DBLP:conf/eurosys/McCunePPRI08} to see how these can be used as an alternative to dedicated hardware based TEEs.  When deployed as TEE environments they generally rely on a Virtual Machine Manager (VMM) to provide the marshaling of access to the resources.  This poses a security concern as the VMM is considered to be part of the Trusted Computing Base (TCB) of such a system, thus increasing the attack surface.
\fi

\subsection{Standardizing TEE Functionality}
\label{subsec:standards}

\begin{figure}[h]
\centering
\includegraphics[width=3.1in]{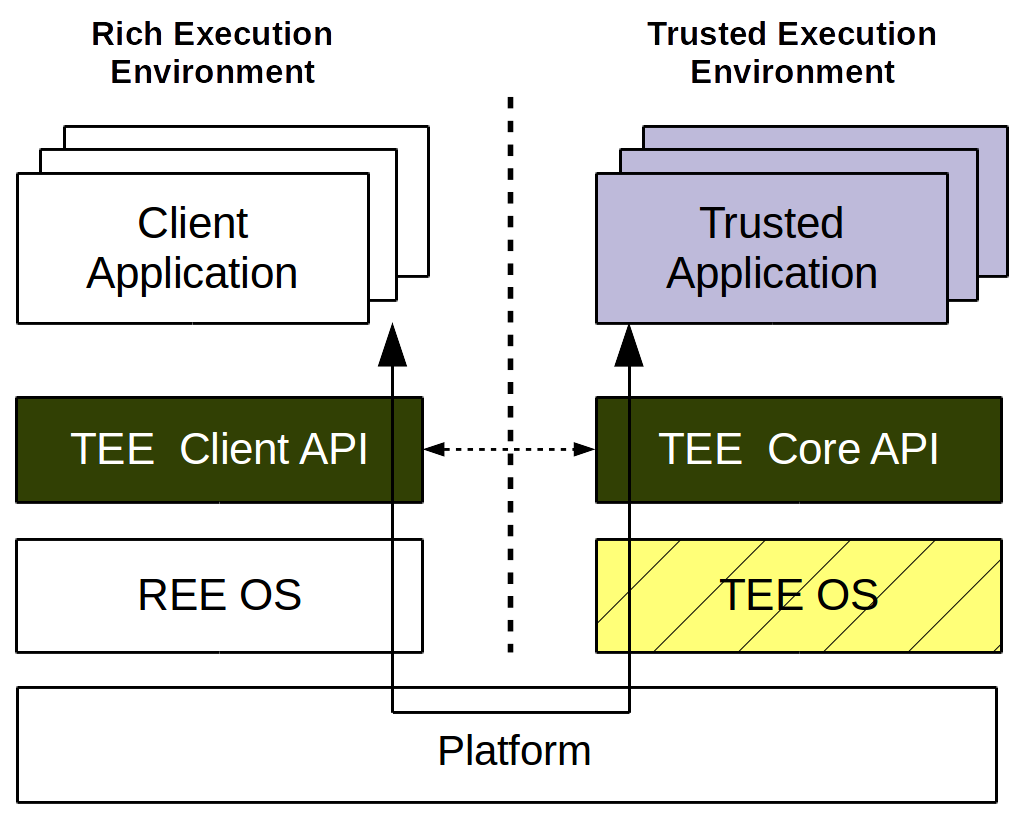}
\caption{The main APIs specified by GlobalPlatform\protect~\cite{web:GP_TEE_SA}}
\label{fig:GP}
\end{figure}

The landscape for TEEs has been very diverse, with a variety of different architectural options from multiple manufacturers. 
\ifabridged
\footnote{For a detailed description of different TEE architectural options, we refer the reader to the full version of this paper~\cite{MDNA15}.}
\fi
Even platforms using the same type of TEE are often not inter-operable. For example, an application written for one TrustZone-based platform will generally not run on a different TrustZone-based platform. They may be using a different TEE OSs or different REE OS drivers. On the other hand, developers and others who are higher up in the software ecosystem are less concerned with intricacies of low-level software or hardware but more concerned with their ability to use the capabilities of TEEs easily and across different platforms. This calls for standardization.
%With all of these varying approaches to TEE design and the different vendors involved we can recognize the need for some form of interoperability or standardization.  The higher up in the software ecosystem, e.g. the operator or the app developer, the less concerned they are with hardware variations in the CPU or SoC and more concerned with the operating system services provided by SDKs that they can leverage to provide additional services to their users and customers.  If an application is written for a TEE from one vendor it generally can not be deployed on a device from another vendor even if they are running on a similar core.  E.g. just because a platform supports TrustZone it does not mean that the implementation or services supported by it will be the same across OEMs.  As a rule of thumb they will not be, and this greatly affects the product portfolios that can be offered because of the different combinations of device vendors and OS that must be supported. 
%%
%% NA Note: reworded

%\subsubsection{GlobalPlatform}
One initiative in TEE standardization has been undertaken by GlobalPlatform (GP)~\cite{web:GP}, which ``is a cross industry, non-profit association which identifies, develops and publishes specifications that promote the secure and inter-operable deployment and management of multiple applications on secure chip technology''~\cite{web:GP_ABOUT}.  GP offers specifications in three areas: smartcards, back-end support systems and devices.  In this paper, we are concerned with specifications from the device working group related to the APIs for TEEs.

Figure~\ref{fig:GP} shows the primary interfaces standardized by GP. 
The \emph{TEE Core API} provides an extensive set of features such as a crypto API and secure storage that can be used to implement a TA, for example a DRM decoder.
The \emph{TEE Client API} is a very generic and thin layer consisting of a small number of functions and definitions that allow the transfer of data back and forth from the REE to a TA.  A CA, for example a DRM player, will implement all complex but non-critical functionality by itself, but use the Client API to invoke the corresponding TA, such as the DRM decoder.
Between the TEE Client API running on the REE and TEE Core API running on the TEE we have an effective Remote Procedure Call (RPC) mechanism where a process running in the REE can invoke tasks in the TEE.

These standardization efforts in GlobalPlatform could resolve the issue of inter-operable TEEs. However, they do not remove the obstacle in gaining access to the requisite hardware nor does simplify the task of developing and testing TAs.

% Design
\section{\openteesectitle}
\label{sec:opentee}
In order to pave the way for the widespread use of TEE functionality by developers and researchers we propose an architecture and a software development kit (SDK) that implements this as a framework atop a set of tools that are familiar to the developer, thus removing the need for specialized hardware and the overheads that it incurs.

\subsection{Motivation}
\label{subsec:motivation}

To explain our motivations, we now revisit the difficulties in developing TEE applications that we alluded to in Section~\ref{sec:intro}:
\paragraph{Enable developer access to TEE functionality}
For a variety of reasons, access to TEEs is generally restricted to developers working for chip manufacturers and to the original equipment manufacturers (OEMs) that make devices based on these chips. Usually, the technology is proprietary and easily deployable SDKs are not available. Furthermore, TEEs may not have a security architecture within them to safely allow complete outsider developers access to TEEs. However, there have been attempts to address this problem~\cite{KostiainenEAR09}.

\paragraph{Provide a fast and efficient prototyping environment}
The most common methods of debugging TAs are to either use expensive JTAG\footnotemark debugging or resort to primitive ``print tracing'' by inserting diagnostic output in the source code. The former generally allows for detailed instruction level debugging. However, the costs associated with these debuggers can be prohibitively expensive, and the setup complex.  Print tracing as a debugging technique is cumbersome and clutters up the source code even to locate the source of a problem. Another concern encountered by TEE developers is that if a TA running on actual device hardware crashes, a hard reset of the device maybe required to recover, thereby significantly increasing the time and effort of debugging.
\footnotetext{Joint Test Action Group standard addresses debugging of integrated circuits}

\paragraph{Promote research into TEE Services}
Ways to isolate TEEs from REEs are reasonably well understood \ifnotabridged{as we saw in Section~\ref{sec:background}}\fi. What is less well understood are the types of services that could benefit from using TEEs.  As the app store model has proven, given an opportunity, the developer community at large is capable of pushing the boundaries and exploring new and novel ways to use technology. Making it possible for researchers to easily develop TAs could trigger the development of novel and innovative applications.

\paragraph{Promote community involvement}
The pre-requisite for involving the developer community and researchers at large is to allow them access to a freely and easily available development environment, SDK and a platform with which to experiment.  The financial and technical aspects of making hardware TEEs available for development on a large scale motivates the need for a software framework for TA development which is not bound to any particular hardware or vendor.

%%
%% NA Note: this text is repetitive and jumps the gun. I have rephrased to make only the point about cost
%%
% To get the community involved, application developers need development environments, SDKs and a platform with which to experiment.  Although the GP standard simplifies conceptualizing a TEE and the functionality that can be offered by it, there are a number of obstacles that the developer must first overcome.  The hardware is complex and expensive to design and manufacture, by its nature it is complicated to deploy, test with and there are no standard tools with which to work.

% Safely exposing TEE functionality to app developers will enable them to innovate these novel approaches to improve  the security and privacy of their applications.  Exposing TEE technology to a wider audience in no way guarantees that all security threats will disappear, however, a community can provide more varied research and ideas than a small group ever could.  This is one of the few ways that we can start to realize the potential of a TEE.  The financial and technical aspects of accessing a hardware TEE make this infeasible so we propose the creation of a TEE framework, that is not bound to any hardware or any particular vendor, yet tries to conform to one standardization effort, for which we have chosen GP.

\subsection{Requirements}
\label{subsec:requirements}
Motivated by the above discussion, our aim is to develop an SDK and framework that allows for the development and testing of standard-compliant TEE applications. The framework should allow development of GP-compliant CA and TA functionality without having to rely on any particular hardware support.  We intend this to be a fast prototyping and development environment that also provides a platform from which to conduct further research into TEE functionality.  Our fundamental design principle is that it should require as little configuration and maintenance as possible, allowing the developer to focus on the task at hand.

We identify the following criteria by which we can measure our TEE framework's usefulness and hence its potential success in addressing the issues that motivated it:
%These decisions have further influenced the design choices that were made during the implementation, what follows are a number of the key points that we set out to address.

\noindent\textbf{Compliance}: Our framework should comply with GP's main interfaces, the Client and Core APIs.

\noindent\textbf{Hardware-independence}: As a software based solution our framework should not be dependent on a particular TEE hardware environment. It should also not be dependent on any particular hardware for the development system itself.

%%
%% NA Note: since we don't know how to evaluate this requirement, we
%% will leave it out for now.
%\noindent\textbf{API extensibility}: It must be flexible to allow easy modification of the GP API if and when an updated version of the API is made available.

\noindent\textbf{Reasonable Performance}: To be readily deployed, our framework must not suffer from code bloat that adds to the on-disk footprint nor to the memory consumption required to run it.  In addition the start-up and restart times of the environment, especially that of the CAs and TAs should not be excessive.  
%One of the perceived benefits of our framework is it's ability to support fast prototyping and as such any time penalty that is incurred will diminish its usability and the satisfaction of using it.

\noindent\textbf{Ease-of-use}: The solution should be easily deployed and configured. It should use tools that are widely available making it more attractive (e.g., there should be no need for extra package/tool configuration on the development system).

We now describe our design and implementation of such a software framework which we call \opentee.

\subsection{Architecture}
\label{subsec:arhictectire}
\begin{figure}[h]
\centering
\includegraphics[width=3.5in]{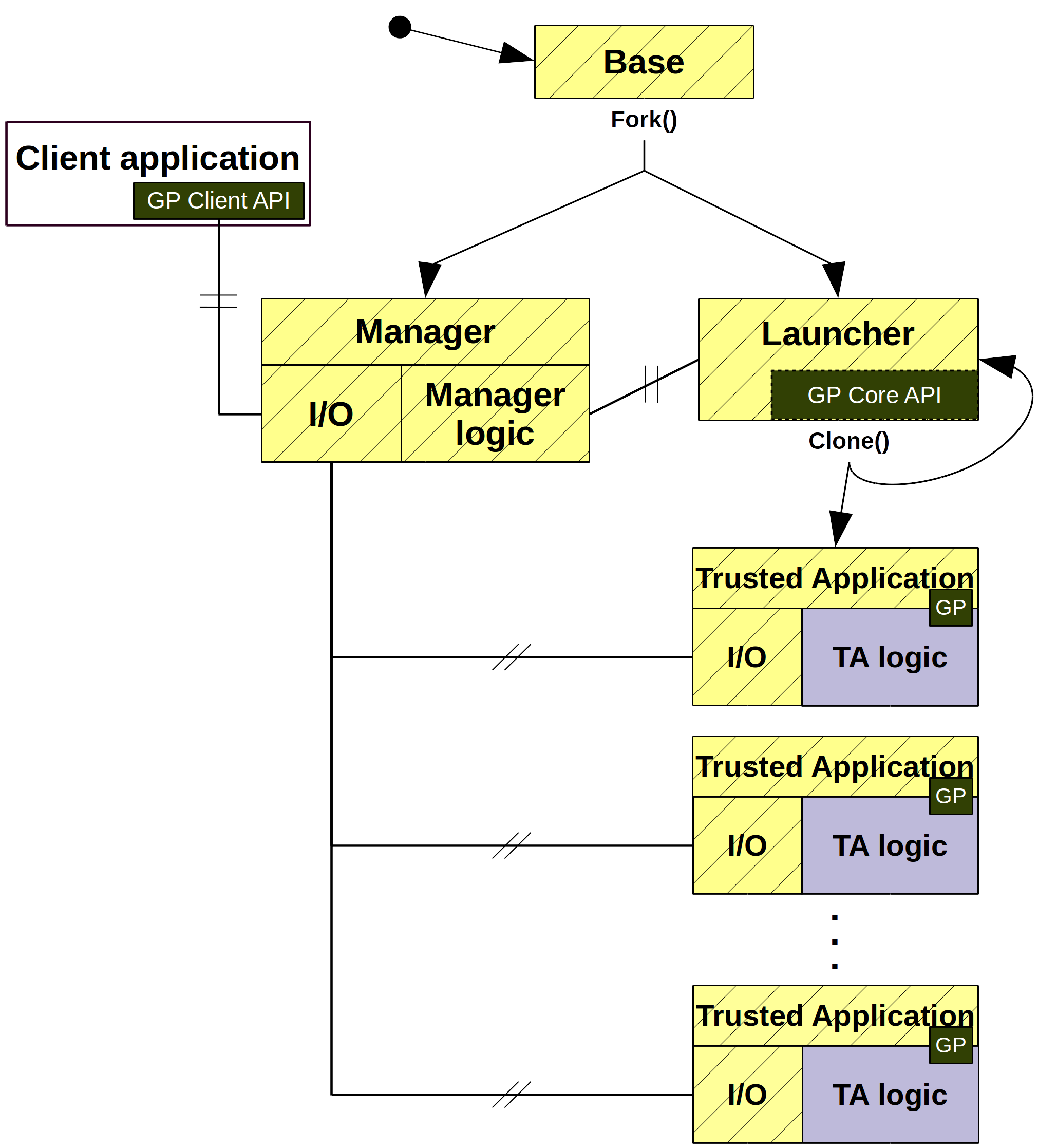}
\caption{\opentee architecture}
\label{fig:opentee}
\end{figure}

We begin with an overview of the structure of the \opentee environment. Figure~\ref{fig:opentee} identifies the main components and their relationships. The color code used in Figure~\ref{fig:opentee} is the same as that used for Figure~\ref{fig:GP} to make the correspondence between the \opentee implementation architecture and the GP conceptual architecture is clear. We describe each component in detail below.
%helps to highlight the relationships between the various parts of the environment e.g. what is the purpose of the process and how it affects the \opentee ecosystem. The presented \opentee solution is based on validation criteria presented in the previous chapter.   

\paragraph{Base}
\opentee is designed to function as a daemon process in user space. It starts executing Base, a process that encapsulates the TEE functionality as a whole.  Base is responsible for loading the configuration and preparing the common parts of the system.  Once initialized the Base will fork to create two independent but related processes. One process becomes Manager and the other, Launcher which serves as a prototype for TAs.

\paragraph{Manager}
Manager can be visualized as \opentee's ``operating system.'' Its main responsibilities are: managing connections between applications, monitoring TA state, providing secure storage for a TA and controlling shared memory regions for the connected applications. Centralizing this functionality into a control process can also be seen as a wrapper abstracting the running environment (e.g. GNU/Linux) and reconciling it with the requirements imposed by the GP TEE standards. GP requirements and the host environment's functionality are not always aligned. For example, GP requirements stipulate that if a TA/CA process crashes unexpectedly, all shared resources of the connected processes must be released. In a typical running environment, this requires additional steps beyond just terminating the process. For example all shared memory must be unregistered -- this needs to be a distinct action from normal process termination.

\paragraph{Launcher}
The sole purpose of Launcher is to create new TA processes efficiently. When it is first created, Launcher will load a shared library implementing the TEE Core API and will wait for further commands from Manager. Manager will signal Launcher when there is a need to launch a new TA (for example, when there is a request from a CA). Upon receiving the signal, Launcher will clone itself. The clone will then load the shared library corresponding to the requested TA. The design of Launcher follows the ``zygote'' design pattern (such as that used in Android~\cite{web:zygote}) of pre-loading common components. This is intended to improve the perceived performance of starting a new TA in \opentee because shared libraries and configurations common to all TAs are pre-loaded into Launcher, the time required to start and configure the new process is minimal. A newly created TA process is then re-parented onto Manager so that it is possible for it to control the TA (so that, for example, it can enforce the type of GP requirements discussed in the paragraph above).

\paragraph{Trusted Application Processes}
The architecture of the TA processes is inspired by the multi-process architecture utilized in the Chromium Project~\cite{web:Chromium-multi-arch}. Each process has been divided into two threads\footnote{The architecture of Manager follows the same division}. The first handles Inter-Process Communication (IPC) and the second is the working thread, referred to respectively as the IO and TA Logic threads. This architectural model enables the process to be interrupted without halting it, as occurs when changing status flags and adding new tasks to the task queue. Additional benefits of this model are that it allows greater separation and abstraction of the TA functionality from the \opentee framework.

\paragraph{GP TEE APIs}
The TEE Client API and TEE Core API are implemented as shared libraries in order to reduce code and memory consumption. 
%In addition, by loading the Core API into Launcher when it is created, will help to reduce the startup times of the TA process removing the need to load it for every TA; this is one of the key benefits of the ``zygote'' design.

\paragraph{IPC}
\opentee implements a communication protocol on top of Unix domain sockets and inter-process signals as the means to both control the system and transfer the messages between the CA and TA.

\subsection{Implementation and Tooling}
\label{subsec:implementation}
\paragraph{Utilizing existing functionality} To meet the hardware-independence requirement, we do not emulate specific TEE hardware with software based emulators, such as QEMU~\cite{web:QEMU}. Instead we rely on existing technologies and the services offered by the mainstream OS of \opentee's running environment rather than developing a new TEE OS to deploy the GP APIs in.  In addition we reuse software from existing open source projects, such as the OpenSSL crypto library and the GNU tool suite, thereby reducing the amount of time required to develop and test the \opentee framework.

This also contributes towards meeting the ease of use requirement in that developers can easily set up \opentee and start developing TAs using a set of familiar tools, editors, IDEs, compilers and debuggers. For example, a developer utilizing \opentee can connect to a TA process with a cheap reliable software debugger such as GDB~\cite{web:GNU_GDB} for detailed debugging tasks like stepping through the code, inspecting variables and registers etc.

\paragraph{Development process} The intended user base for \opentee consists of seasoned developers. To ensure viability in such a demanding user base, we adopted a rigorous development process for \opentee so that the end result will be perceived as robust and usable. \opentee is developed as an open source project and as such there are a number of powerful tools that are freely available for this type of project. As previously mentioned GitHub is used for hosting the code. GerritHub\footnote{http://gerrithub.io/} is used for performing peer-review of all code before it is submitted to the code base.  In addition to the manual review process we leverage the power of Coverity\footnote{https://scan.coverity.com/projects/3441} to perform in depth static analysis scans. This enforces secure coding practices and helps to find potential functional bugs that may have been missed during the manual code review. In addition, we have deployed a continuous integration (CI) server running Jenkins\footnote{http://jenkins-ci.org/}, which we have connected to GerritHub.  Its main task is to perform a number of ``smoke tests''\footnote{A suite of tests intended to ensure that the basic functionality of a system are intact.} on the new patches. These tests ensure that the patches conform to the coding guidelines, build successfully and that the basic system is usable after the patches are applied.

\paragraph{\opentee in Use} Being designed as an open source framework upon which to build and test features that will utilize a TEE, \opentee has been implemented to be as inconspicuous as possible. The complexity of the system is hidden from the users of \opentee.  They are presented with an SDK that exposes the Client and Core APIs without being required to have a deep understanding of how the overall framework works, thereby allowing them to focus on the development of their own TAs. However, \opentee is already being extended by the community. The ongoing implementation of the GP Trusted UI specification is an example.

% Validation
\section{Evaluation}
\label{sec:eval}

%%
%% NA Note: what has this got to do with validation?
%Being an active open source project it is constantly evolving and more features are being added and address based upon feedback that has been received.
We now return to the requirements from Section~\ref{sec:opentee} and evaluate how well \opentee meets those requirements.

\subsection{Compliance}

Every effort has been made to comply with the GP standard. Whenever this has not been feasible, due to time constraints or in the interest of providing a platform upon which to build, the deviation has been documented and a debug message is logged to inform the user of the non-compliance. The Client API is fully implemented. The Core API implementation has 100\% function coverage, however, the cryptographic algorithm coverage is currently 80\% due to the use of existing libraries that do not support the remaining algorithms.
A compliance test suite is commercially available from GP. There is no information about how well existing TEE hardware conform to the GP specifications. Based on our experience in implementing version 1.0.26 of the GP Core API, we provided detailed feedback, including on errors and ambiguities in the specification, to GlobalPlatform in response to their solicitation of public comments. Several items in our feedback have been addressed in the released version 1.1.
% However, as an open source project we lack the funding to purchase the use of this tool.  Information related to other implementations of the GP specification are scarce and most of the information related to it is proprietary and therefore it is not know if some of the commercial TEE implementations are fully compliant. If this were known, an idealistic validation of our implementation would be executing CAs and TAs we have created in an independently implemented TEE to compare the results.

\subsection{Hardware-independence}

By following the GP standard and not emulating any specific TEE hardware, \opentee is independent of TEE hardware. TAs developed with \opentee can be compiled to any target TEE hardware architecture. We have verified~\cite{ekberg15} that a non-trivial a TA developed using \opentee (284 LoC, 19 Internal Core API invocations (9 unique functions), 6 invokable TA commands) has been succesfully compiled and run on a
hardware TEE based on ARM TrustZone running the Trustonic <t-base environment~\cite{web:trustonic_t_dev}.

%The implication of this is seen in the performance measurements of the TAs and CAs that are being developed.  
\opentee can provide coverage reports to help highlight hot-spots in the code, generate call graphs etc. The GP Internal Core API includes memory management primitives and allows configuration parameters (such as \texttt{gpd.ta.dataSize} and \texttt{gpd.ta.stackSize}) to indicate how much heap and stack memory is available to a TA. A developer can use these parameters to configure \opentee to reflect the memory restrictions of a target hardware TEE environment.
However, as the actual TEE is potentially running a different environment than that offered by \opentee\ -- possibly utilizing hardware based cryptographic accelerators, potentially having a different CPU, with different clock speed and throughput characteristics – it will result in different timing characteristics.  In this sense, as with all virtual environments, \opentee can not fully replace the actual hardware environment for the final stages of the development cycle. Instead developers using \opentee can gain confidence that the hardware-independent parts of their trusted applications have been optimally implemented by making judicious use of coverage reports and other generic analysis techniques. Any hardware-specific optimization, such as performance tuning, naturally needs to be done on the target hardware environment.

\opentee has been deployed and used on various development environments ranging from servers to desktops and laptops.
%\footnote{Lenovo X1 carbon}\footnote{MacBook pro}\footnote{Samsung XE303C12, running Linux} 
We have tested \opentee on both ARM and x86 architectures. \opentee requires Linux but has been run successfully on virtual machines hosted on other OSs.
%Having chosen not to emulate existing hardware to create the framework helps to ensure that the TAs created using it are portable as it is harder to create machine dependent code.

%%
%% NA Note
%% This is not a robust defense. So we will leave it out for now at least.

% \subsection{API extensibility}
  
% The implementation started with GP Core API version 1.0 and was almost complete when GP announced a subsequent point release of this specification. This afforded us the opportunity to validate the API extensibility requirement we had set. However small the delta between 1.0 and 1.1 was, it nevertheless validated our design choices. The estimated effort required for this transition could be perceived as subjective, but in our in opinion the effort was small and the transition was relatively straightforward.

\subsection{Footprints and Performance}

To evaluate our performance we deployed \opentee on a desktop machine (Intel i7-2600 CPU with 8GB RAM) running 64-bit Ubuntu 14.04. All performance tests were run 40 times while the machine was under normal load e.g. having editors and browsers open.

\subsubsection{Disk and Memory consumption}
\opentee is written in ANSI C with a total of 12423 lines\footnote{gathered using sloccount: http://www.dwheeler.com/sloccount/} spread over 78 source and header files. Table~\ref{tab:sizes} shows the total size of the framework and highlights two libraries from the framework that are of most interest to developers, being \texttt{ libInternalApi.so} against which the TAs are linked and \texttt{libtee.so} against which CAs are linked. As is standard on operating systems that support shared libraries the ``Text'' section, containing the programs code, can be shared among the different processes that link against it. The ``Data'' and ``BSS'' respectively refer to the initialized and uninitialized data parts of the library that can be shared in a Copy-On-Write (COW) basis. As the table highlights the vast majority of the libraries' size can be shared, thus reducing the required footprint.

\begin{table}
\centering
\caption{Binary sizes (bytes)}
\label{tab:sizes}
\begin{tabular}{l c c c c}
  & Text & Data & BSS & overall \\
  \hline
  libInternalApi.so & 117448 & 2248 & 160 & 119856 \\
  \rowcolor{black!15} \cellcolor{black!15}libtee.so & 18617 & 880 & 152 & 19649 \\
  Total Framework & 224948 & 7760 & 1664 & 234372 \\
\hline
\end{tabular}
\end{table}

\begin{table}
%\begin{minipage}{\columnwidth}
\begin{center}
\caption{Memory usage (KB)}
\label{tab:memoryUsage}
\begin{tabular}{l c c c c}
  & RSS & Shared & Private & PSS \\
  \hline
  \llap{\smash{\rotatebox[origin=c]{90}{%
    \hspace*{-12pt} % Vertical adjustment
    \tiny{no TA}}}
   \hspace*{10pt}} % Horizontal adjustment
  Manager  & 1024 &  764 & 260 & 305 \\\rowcolor{black!15}\cellcolor{black!15}
  Launcher & 1624 & 1232 & 392 & 558 \\
  \hline
  \hline
  \llap{\smash{\rotatebox[origin=c]{90}{%
    \hspace*{-24pt} % Vertical adjustment
    \tiny{one TA}}}
    \hspace*{10pt}} % Horizontal adjustment
  Manager  & 1112 &  832 & 280 & 316 \\\rowcolor{black!15}\cellcolor{black!15}
  Launcher & 1648 & 1548 & 100 & 397 \\
  \hspace{3pt}Test TA1\tablefootnote{ta\_conn\_test\_app \textasciitilde100 lines of C} & 1072 & 932 & 140 & 308 \\
  \hline
  \hline
  \llap{\smash{\rotatebox[origin=c]{90}{%
    \hspace*{-32pt} % Vertical adjustment
    \tiny{two TAs}}}
    \hspace*{10pt}} % Horizontal adjustment
  Manager  & 1116 &  832 & 284 & 319 \\\rowcolor{black!15}\cellcolor{black!15}
  Launcher & 1648 & 1548 & 100 & 337 \\
  \hspace{3pt}Test TA1 & 1072 &  944 & 128 & 245 \\\rowcolor{black!15}\cellcolor{black!15}
  Test TA2\tablefootnote{example\_digest\_ta \textasciitilde140 lines of C} & 1236 & 1068 & 168 & 299 \\
  \hline
\end{tabular}
\end{center}
%\end{minipage}
\end{table}

Table~\ref{tab:memoryUsage} shows the memory consumption of the running process under three different scenarios.  The first shows the memory consumption of Manager and Launcher immediately after they have been launched, i.e. before any TAs have been launched. The next section shows how the memory consumption increases when one TA is launched while the last sections shows the situation when two TAs are running simultaneously.

\noindent\textbf{RSS (Resident Set Size)} shows how much memory has been allocated for a process, this includes all memory that a process shares with other processes. As such it is very naive measurement of a processes memory impact.

\noindent\textbf{Shared} is the memory that a process shares with other processes, i.e. through the use of shared libraries.

\noindent\textbf{Private} is the memory that is private to a process and will be returned to the system when the process terminates, however, Copy-On-Write semantics after a process fork may complicate this calculation.  The Private pages may actually be shared until one or the other of the processes tries to write to the page, at which time it will be given its own copy of the Private page.

\noindent\textbf{PSS (Proportional Set Size)} is a realistic indicator of the actual memory footprint of a process.  It is calculated as the sum of the Private memory used by a process and the average Shared memory use per process. 
%(Shared memory / number of processes sharing it) 
E.g., if a process has 100KB of Private memory and 1000KB of memory shared with 10 processes, its impact on system memory is 200KB\footnote{100KB + (1000KB / 10) = 200KB}. Taking the example of Launcher between runs 2 and 3 we see that while the RSS, Shared and Private memory usage stay constant the PSS decreases as more pages are shared with the new TA.

Overall, we can conclude that (a) the memory footprint of \opentee is low and (b) the extensive use of shared libraries implies that the marginal memory cost of launching a new TA is small, as shown by the PSS figures.

\subsubsection{Build and Run performance}
One of the driving requirements of \opentee is the need to have short build and deploy cycles to help reduce the overall development effort. Table~\ref{tab:performance} highlights that \opentee does not pose a significant overhead to the developer, taking an average of just 147 ms to perform an incremental build of a TA. The time required for an incremental build was comparable to that of a clean build, falling within the standard deviation of the former, this can be attributed to the source code being confined to a single C file. Comparative results are not available for deployed hardware-based development environments. However, considering that a full reset of the target device and the subsequent boot of its OS may be required before the CA can be launched, \opentee's performance is likely to be perceived as being superior.

\begin{table}
\centering
\caption{Average build and execute times of a TA, including standard deviations}
\label{tab:performance}
\begin{tabular}{l c}
  & Time \\
  \hline
  Build & 147 ms $\pm$ 10.95 \\
  \rowcolor{black!15} \cellcolor{black!15}Execute & 430.5 $\mu$s $\pm$ 32.6 \\
\hline
\end{tabular}
\end{table}

\subsection{Ease of use}
\label{user-study}

Determining whether \opentee eases the burden of TA development is particularly
challenging because, until now, TA development has been limited to a very small
set of developers. Fortunately, we were able to recruit several experienced TA
developers from multiple organizations to participate in a user
study.
\ifabridged
Fourteen people participated in the study. Eleven had prior experience
developing/debugging TAs. We used a pre-study and a post-study
Standard System Usability Scale
(SUS)~\cite{bookchapter:sus,jour:BKM08} questionnaire to
elicit the participants' estimates of the ease of use in developing
TAs. They were asked to complete the pre-study SUS questionnaire with
their current TA software development environment in mind and the
post-study SUS questionnaire with \opentee in mind.
See the full version of this paper~\cite{MDNA15} for a complete
description of the user study.
\else
Our user study was conducted as follows.

\noindent\textbf{Participants}: Fourteen people participated in the
study. All had prior software experience (between 3 and 33 years,
$M=13$, $SD=8.2$). Eleven had prior experience developing/debugging
TAs (between \textonehalf\ and 15 years, $M = 5.1$, $SD = 4.2$).

\noindent\textbf{Materials}: The Standard System Usability Scale
(SUS)~\cite{bookchapter:sus,jour:BKM08} questionnaire was used to
elicit the participants' estimates of the ease of use in developing
TAs. We used a pre-study and a post-study questionnaire. In addition
to demographic information, the pre-study questionnaire included
free-form questions about the current software development environment
(if any) they use for TA development.  It also contained a SUS
questionnaire which the participants were asked to complete with their
current TA development environment in mind. This was completed only
by those participants who had prior TA development experience.
\fi
The material for the user study\footnotemark task was a sample CA/TA
pair, provided as part of the \opentee source tree. A software flaw
had been introduced to the TA, which, when executed, would result in a
segmentation fault and subsequent premature termination of the TA. The
CA was free of error and was only used to interact with the TA running
in \opentee.
\ifnotabridged
Code complexity metrics for the TA used in the user study are shown in
Table~\ref{tbl:user-study-ta-metrics}.
\fi

\ifabridged
\else
The post-study questionnaire consisted of a SUS form which the participants
were asked to complete with \opentee in mind. The questionnaire also had open
ended questions about specific difficulties they face in TA development.
\fi

\footnotetext{
The material can be found at
\ifanonymous
\url{https://drive.google.com/folderview?id=0B6MDl5z3gmhedmdLVVJpaXREeXM}
\else
\url{http://open-tee.github.io/userstudy/}
\fi}

\noindent\textbf{Procedure}: The user study was conducted in three
steps. In the first step, participants were first asked to complete
the pre-study questionnaire. After this they were pointed to a web
page containing brief instructions on how to install and use \opentee.
In the second step, once the participants completed the tutorial they
were told about the flawed TA. They were tasked to identify the reason
for the TA malfunction using \opentee and correct the flaw in the TA.
Finally, in the third step, after the participants had completed the
debugging exercise, they were asked to complete the post-study
questionnaire.

\begin{figure}
\includegraphics[width=\textwidth/2]{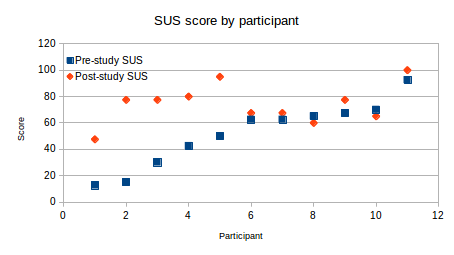}
\caption{Pre- and post-study SUS score (for participants with prior TA
development experience)}
\label{fig:user-study-scores}
\end{figure}

\ifnotabridged
\begin{table}
\centering
\caption{Complexity metrics for TA used in user study}
\label{tbl:user-study-ta-metrics}
\begin{tabular}{l r}
  \hline
  Total Physical Source Lines of Code        & 284 \\\rowcolor{black!15} \cellcolor{black!15}
  Total Number of Invokable TA Commands      & 6   \\
  Total Number of Internal Core API Calls    & 16  \\\rowcolor{black!15} \cellcolor{black!15}
  Number of Distinct Internal Core API Calls & 9   \\
  \hline
\end{tabular}
\end{table}
\fi

\ifabridged
% Omit experience figure in abridged version
\else
\begin{figure}
\includegraphics[width=\textwidth/2]{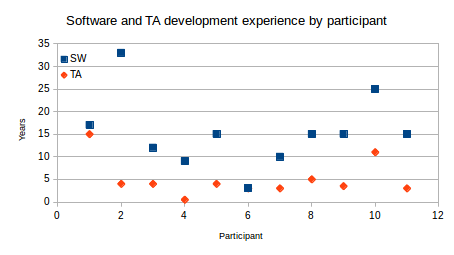}
\caption{General software and TA development experience by participant (for
participants with prior TA development experience)}
\label{fig:user-study-experience}
\end{figure}
\fi

\begin{table}
\centering
\caption{Mean, standard deviation and median for the pre- and post-study SUS scores}
\label{tbl:user-study-scores}
\begin{tabular}{l c c c}
  & Mean & Std.dev. & Median \\
  \hline
  Pre-study SUS & 51.82 & 24.70 & 62.50 \\
  \rowcolor{black!15} \cellcolor{black!15}Post-study SUS & 74.09 &
  15.01 & 77.50 \\
  Post-study SUS \tiny(all participants) & 69.92 & 18.09 & 68.75 \\
  \hline
\end{tabular}
\end{table}

\noindent\textbf{Results}: 
The mean, standard deviation and median of the SUS scores for all
participants, including those without prior TA development experience
are shown in Table~\ref{tbl:user-study-scores}. With both sets of
participants, the post-study questionnaire yields a mean score above
68, which is considered the threshold value for an above average SUS
score, indicating an acceptable level of usability in \opentee.

Figure~\ref{fig:user-study-scores} shows the scores reported both
before and after the use of \opentee by participants with prior TA
development experience. Nine out of the eleven participants (82\%) rated
\opentee higher than the development environment they are using
currently. This suggests that the perceived usability of \opentee is
higher than that of the current tools used by the experienced TA
developers. In five cases (46 \%), the difference in SUS scores was 35
or more. In the remaining six cases, the difference in SUS scores was
10 or less. A Wilcoxon signed-rank test showed that the difference in
SUS scores is statistically significant ($z=-2.50$, $p<.05$, $r=-0.53$).

\ifabridged
In the full version of this paper~\cite{MDNA15} we present and discuss
extracts from participant responses to open-ended questions about TA development.
\else
The difference in SUS scores divides the participants into two
distinct groups. The five participants for whom the difference was 35
or more had SUS scores below 60 in their pre-study questionnaire. The
remaining six for whom the difference was 10 or less had pre-study SUS
scores over 60. A natural question is whether we can discern any
other difference between the two groups that might explain the
difference in SUS scores. One possible explanation was that
experienced software developers were comfortable with their current
tools and hence did not perceive \opentee as being easier to use. If
this explanation is correct then one can hypothesize that developers
with many years of general software or TA development experience will
rate their current development tools higher than their counterparts
with fewer years of experience would. However, a Spearman's rho
correlation test indicated no significant correlation between the
years of general software development experience and the SUS score in
the pre-study questionnaire ($r_{s}=-.042$, $p>.05$), nor between the
years of TA development experience, and the SUS score in the pre-study
questionnaire ($r_{s}=-.204$, $p>.05$). 
%\ifabridged
% Omit reference to experience figure in abridged version
%\else 
Figure~\ref{fig:user-study-experience} shows the software
development experience (both general and TA) reported by each
participant whose SUS scores are shown in
Figure~\ref{fig:user-study-scores}. 
%\fi

A majority of the experienced TA developers (7 out of 11, 64 \%),
reported using hardware tools for debugging TAs under
development. Four (36 \%) used
Lauterbach\footnote{\url{http://www.lauterbach.com/}} hardware
assisted debug tools. Three (27 \%) used other development boards such
as Arndale\footnote{\url{www.arndaleboard.org/}},
Fido\footnote{\url{http://www.liewenthal.ee/projects/fido/}} or
DS-5\footnote{\url{http://ds.arm.com/}} or actual mobile
devices. Participant responses highlighted different types of difficulties in
debugging TAs using only hardware:
\begin{itemize}
\itemsep0em
\item workflow slowdown due to the need to (cross) compile, load and
  execute TAs on separate hardware (\textit{``slow execution (flash,
    download, reboot, run)''}, \textit{``debugging TA is slow, you
    need to cross compile and push binary into target hardware''}),
\item problems due to the hardware itself being under development and
  hence exhibiting flaws, (\textit{``TEE itself might not work without
    problems, because some change have been made''}), and
\item inconvenience caused by the restricted access to prototype
  hardware \textit{``Main difficulty is that you need development
    hardware, which is problematic when working outside the
    office.''}).
\end{itemize}
Six participants (55 \%) reported that their current development
environment does not support interactive debugging. But even the rest,
who used tools like Lauterbach tracing, reported
that they found it easier to resort to print tracing,
whenever they needed to examine values of TA variables.

After having used \opentee, several participants commented
\textit{``debugging is easy''} or \textit{``debugging is fast''} in
the post-study questionnaire. One participant characterized how
\opentee could be integrated into his existing workflow before
cross compiling to target hardware: \textit{``[\opentee] complements
nicely my previous SDE - first preliminary testing with \opentee \&
\texttt{gdb} \& \texttt{OT\_LOG(..)}, and only after that ARM cross
compiler \& FVP emulation''}. The dominant suggestion for
improvement was a desire to see more extensive documentation for
\opentee.
%One recurring suggestion was: 
%\textit{``A template for both [CA and TA]
%and some scripts to start developing would be great''}

Given the sample size, the results should be taken as indicative
rather than definitive. However, it is reasonable to conclude that
\opentee has the potential to improve the ease of use of developing
TEE applications.
\fi

% Related work
\section{Related Work} 
\label{sec:relatedwork}

Ekberg et al.~\cite{EkbergKA14} list several reasons for the underutilization of TEEs in devices: e.g., lack of standard APIs and easily available SDKs and lack of trust between the different stakeholders, with OEMs being unwilling to open up their security environments to third parties.  In this section we will review a number of initiatives that have been undertaken to address some of these issues and compare these efforts to \opentee.

ObC~\cite{KostiainenEAR09} was one of the first attempts to address the problem of opening the TEE to third party developers by challenging the prevailing opinion that a credential system must be centralized and closed. ObC predates many standardization efforts and as such defines a proprietary mechanism by which to enable the CA/TA communication and synchronization while leveraging the TrustZone architecture to enforce the security. On the other hand our work aims to promote standards adoption in order to proliferate TEE research and deployment.

Muthu~\cite{thesis:QEMU} analyzes extending QEMU to support TrustZone, the feasibility of such a solution, and tries to determine if it would be beneficial to the developer community. Winter et al.~\cite{WinterWPT11} go one step further and implement a TrustZone emulator as an open source project. However, we were not able to find the source code. \opentee addresses the issue of virtualizing the TEE, however, in contrast we are not tied to the emulation of a specific TEE implementation. One issue with developing an emulator for the TEE is that it still lacks an operating system to run. In section ~\ref{subsec:standards} we highlight the lack of a standardized OS even among the different TrustZone implementations.

To this end there have been a number of efforts to create an OS that is suitable to be deployed in TrustZone~\cite{web:NVIDIA_TLK}~\cite{web:linaro_op_tee}~\cite{web:T6}. All of these are open source solutions which are released under various licenses (Table ~\ref{tab:alternatives}). In addition to providing an operating system for the TEE both OP-TEE~\cite{web:linaro_op_tee} and T6~\cite{web:T6} choose to rely on GP as their RPC mechanism between the REE and TEE. TLK~\cite{web:NVIDIA_TLK} on the other hand chooses to provide a proprietary communication mechanism.

Sierraware's Open Virtualization~\cite{web:openVirtua} provides a dual-licensed OS implementation\footnote{proprietary, GPL} that also supports the GP standards. The commercial products (sierraVisor, SierraTEE) provide extended functionality and not being GPL there is no requirement for any changes to be made  publicly available by the license holders as is required with their open source offering. \opentee is licensed under Apache-V2 giving users the flexibility of an open source license without the strict copyleft requirements.

Trustonic's <t-dev developers program~\cite{web:trustonic_t_dev} was created to support Trustonic partners who have deployed the <t-base TEE implementation. This program provides an SDK, tools and consulting with the aim of easing the development and testing of TEE applications in deployed hardware solutions.

All of the OS based solutions have to be ported to support the various HW environments, increasing the effort of maintaining the OS and reducing the users available options.  Many of them also require that the HW be configured in a developer mode, without this setting it is generally not possible to deploy custom SW to the TEE, for obvious reasons, further restricting the developers options.  \opentee in contrast does not have this HW dependency, thus enabling the users to start developing with the framework once they have cloned the repositories. Based on the references listed above, we conclude that no other project fills the niche of a fast prototyping SDK framework that we have described in this paper.

\begin{table}
\centering
\caption{Comparison of available alternatives to \opentee}
\label{tab:alternatives}
\begin{tabular}{l c c c}
  & Compliance & HW-independence & License \\
  \hline
  \opentee & yes & yes & Apache-V2 \\
  \rowcolor{black!15} \cellcolor{black!15}Open Virtualization~\cite{web:openVirtua} & yes & no & proprietary, GPL \\
  OP-TEE~\cite{web:linaro_op_tee}  & yes & no & BSD-2,BSD-3 \\
  \rowcolor{black!15} \cellcolor{black!15}T6~\cite{web:T6} & ? & no & ? \\
  TLK~\cite{web:NVIDIA_TLK} & no & no & MIT,FreeBSD \\
  \rowcolor{black!15} \cellcolor{black!15}TrustZone Emulator~\cite{WinterWPT11} & ? & no & ? \\
\hline
\end{tabular}
\end{table}

% Conclusion
\section{Discussion and Conclusion}

We have demonstrated that \opentee meets our objective of an easy-to-use, hardware-independent software framework that allows developers to write and debug GP-compliant TEE applications. We made a deliberate decision to open source \opentee under Apache-V2 license~\cite{web:Apache-v2}. The Apache license was selected because it is a recognized open source license and it provides additional flexibility for those wishing to use the framework. All third party components have been carefully selected -- we have used only components that have been properly licensed and do not set any restrictions for future use. This has made it possible for people from outside the research group to contribute to \opentee. Currently a number of extensions are being worked on including support for other GP APIs and supporting Client API bindings in Java (for Android applications).

Although the sample in our user study is small, participants were drawn from several different organizations with track records of TA development. This makes us confident that our user study results are valid. It is very difficult at this time to conduct larger-scale user studies of TA development because the community of TA developers is tiny. Recall that expanding the size of the TA developer base is the very motivation for \opentee in the first place.

We initially intended \opentee as a developer tool. However, an alternative use has become evident in our discussions with service providers. Although use of TEEs can improve the security and usability of their service, not all their clients may have TEE-equipped devices. Yet the service provider would like to present a consistent user experience for their entire client base. A possible approach for them is to ship their application (CA and TA) with \opentee and arrange for the CA to use \opentee if it cannot detect a real hardware TEE on the device. This would allow the service provider to have a common provisioning mechanism and offer a consistent user experience for all their clients. However, once we cast \opentee as a potential fall-back TEE in this manner, we need to address the question of how best to isolate it from the REE in the absence of any hardware support. This is part of our current work.

We re-iterate that \opentee is not intended to emulate any specific TEE hardware. \opentee meets its goal of guaranteeing that trusted applications developed using it will compile and run on any GP-compliant TEE hardware. Hardware-specific aspects, such as performance tuning are outside the scope of \opentee.

Our hope in writing this paper is to make the research community aware of \opentee and encourage researchers to use it and contribute to its development. We also believe that organizations and developers who already develop TA applications will benefit from incorporating \opentee into their development process.

\ifanonymous
% no acks
\else
\noindent\textbf{Acknowledgments}: We thank the participants of the
user study for their time and their useful feedback. Thanks to
Kumaripaba Athukorala for advice on conducting the user study. This
work was supported by the Intel Collaborative Institute for Secure
Computing at Aalto University (\url{http://www.icri-sc.org/}), which
is jointly funded by Intel Corporation and Aalto University.
\fi

\bibliographystyle{IEEEtran}
\bibliography{opentee}

\begin{thebibliography}{10}
\providecommand{\url}[1]{#1}
\csname url@samestyle\endcsname
\providecommand{\newblock}{\relax}
\providecommand{\bibinfo}[2]{#2}
\providecommand{\BIBentrySTDinterwordspacing}{\spaceskip=0pt\relax}
\providecommand{\BIBentryALTinterwordstretchfactor}{4}
\providecommand{\BIBentryALTinterwordspacing}{\spaceskip=\fontdimen2\font plus
\BIBentryALTinterwordstretchfactor\fontdimen3\font minus
  \fontdimen4\font\relax}
\providecommand{\BIBforeignlanguage}[2]{{%
\expandafter\ifx\csname l@#1\endcsname\relax
\typeout{** WARNING: IEEEtran.bst: No hyphenation pattern has been}%
\typeout{** loaded for the language `#1'. Using the pattern for}%
\typeout{** the default language instead.}%
\else
\language=\csname l@#1\endcsname
\fi
#2}}
\providecommand{\BIBdecl}{\relax}
\BIBdecl

\bibitem{EkbergKA14}
\BIBentryALTinterwordspacing
J.~Ekberg, K.~Kostiainen, and N.~Asokan, ``The untapped potential of trusted
  execution environments on mobile devices,'' \emph{{IEEE} Security {\&}
  Privacy}, vol.~12, no.~4, pp. 29--37, 2014. [Online]. Available:
  \url{http://dx.doi.org/10.1109/MSP.2014.38}
\BIBentrySTDinterwordspacing

\bibitem{white:mshield}
J.~Azema and G.~Fayad, ``{M-Shield} mobile security technology,'' 2008, {TI}
  White paper. \url{http://focus.ti.com/pdfs/wtbu/ti_mshield_whitepaper.pdf}.

\bibitem{web:TRUSTZONE}
ARM, ``{ARM} security technology --- {B}uilding a secure system using
  {TrustZone} technology,''
  \url{http://infocenter.arm.com/help/index.jsp?topic=/com.arm.doc.prd29-genc-009492c/index.html},
  April 2009.

\bibitem{web:ARM_TZ}
------, ``Technical reference manual: {ARM} 1176jzf-s (trustzone-enabled
  processor),''
  \url{http://www.arm.com/pdfs/DDI0301D_arm1176jzfs_r0p2_trm.pdf}.

\bibitem{web:SEP}
Intel, ``{SEP} driver,''
  \url{https://git.kernel.org/cgit/linux/kernel/git/stable/linux-stable.git/tree/drivers/staging/sep?id=refs/tags/v3.14.32}.

\bibitem{web:APPLE_SE}
{Apple}, ``{iOS} security,''
  \url{https://www.apple.com/ca/iphone/business/docs/iOS_Security_Feb14.pdf}.

\bibitem{web:SB962}
M.~Leno, ``Senate bill 962, leno. smartphones.''
  \url{http://leginfo.legislature.ca.gov/faces/billNavClient.xhtml?bill_id=201320140SB962}.

\bibitem{Dyer+01}
\BIBentryALTinterwordspacing
J.~G. Dyer, M.~Lindemann, R.~Perez, R.~Sailer, L.~van Doorn, S.~W. Smith, and
  S.~Weingart, ``Building the {IBM} 4758 secure coprocessor,'' \emph{{IEEE}
  Computer}, vol.~34, no.~10, pp. 57--66, 2001. [Online]. Available:
  \url{http://doi.ieeecomputersociety.org/10.1109/2.955100}
\BIBentrySTDinterwordspacing

\bibitem{TPM}
``{Trusted Platform Module (TPM) Specifications},''
  \url{https://www.trustedcomputinggroup.org/specs/TPM/}.

\bibitem{web:GP}
GlobalPlatform, ``Home page.'' \url{http://www.globalplatform.org}.

\bibitem{web:GP_TEE}
------, ``Device specifications for trusted execution environment.''
  \url{http://www.globalplatform.org/specificationsdevice.asp}.

\bibitem{web:GP_TEE_SA}
------, ``{TEE System Architecture},''
  \url{http://www.globalplatform.org/specificationsdevice.asp}.

\bibitem{EkbergDiss}
J.-E. Ekberg, ``Securing software architectures for trusted processor
  environments,'' Doctoral dissertation, Aalto University, May 2013,
  \url{http://urn.fi/URN:ISBN:978-952-60-3632-8}.

\bibitem{SGX13}
\BIBentryALTinterwordspacing
F.~McKeen \emph{et~al.}, ``Innovative instructions and software model for
  isolated execution,'' in \emph{Proceedings of the 2Nd International Workshop
  on Hardware and Architectural Support for Security and Privacy}, ser. HASP
  '13.\hskip 1em plus 0.5em minus 0.4em\relax New York, NY, USA: ACM, 2013, pp.
  10:1--10:1. [Online]. Available:
  \url{http://doi.acm.org/10.1145/2487726.2488368}
\BIBentrySTDinterwordspacing

\bibitem{web:SGX}
Intel, ``Intel software guard extensions (intel sgx),''
  \url{https://software.intel.com/en-us/intel-isa-extensions#pid-19539-1495}.

\bibitem{web:SGX2}
------, ``Software guard extensions programming reference,''
  \url{https://software.intel.com/sites/default/files/329298-001.pdf}.

\bibitem{jour:cheng2013appshield}
Y.~Cheng, X.~Ding, and R.~Deng, ``Appshield: Protecting applications against
  untrusted operating system,'' \emph{Singaport Management University Technical
  Report, SMU-SIS-13}, vol. 101, 2013.

\bibitem{DBLP:conf/eurosys/McCunePPRI08}
\BIBentryALTinterwordspacing
J.~M. McCune, B.~Parno, A.~Perrig, M.~K. Reiter, and H.~Isozaki, ``Flicker: an
  execution infrastructure for {TCB} minimization,'' in \emph{Proceedings of
  the 2008 EuroSys Conference, Glasgow, Scotland, UK, April 1-4, 2008}, 2008,
  pp. 315--328. [Online]. Available:
  \url{http://doi.acm.org/10.1145/1352592.1352625}
\BIBentrySTDinterwordspacing

\bibitem{web:GP_ABOUT}
GlobalPlatform, ``About.'' \url{http://www.globalplatform.org/aboutus.}

\bibitem{KostiainenEAR09}
\BIBentryALTinterwordspacing
K.~Kostiainen, J.~Ekberg, N.~Asokan, and A.~Rantala, ``On-board credentials
  with open provisioning,'' in \emph{Proceedings of the 2009 {ACM} Symposium on
  Information, Computer and Communications Security, {ASIACCS} 2009, Sydney,
  Australia, March 10-12, 2009}, 2009, pp. 104--115. [Online]. Available:
  \url{http://doi.acm.org/10.1145/1533057.1533074}
\BIBentrySTDinterwordspacing

\bibitem{web:zygote}
{Android Open Source Project}, ``Managing your app's memory,''
  \url{https://developer.android.com/training/articles/memory.html}.

\bibitem{web:Chromium-multi-arch}
{The Chromium Projects}, ``Multi-process architecture,''
  \url{http://www.chromium.org/developers/design-documents/multi-process-architecture}.

\bibitem{web:QEMU}
QEMU, ``Open source processor emulator,'' \url{http://wiki.qemu.org/Main_Page}.

\bibitem{web:GNU_GDB}
GNU, ``{GDB}: The {GNU} project debugger,''
  \url{http://www.gnu.org/software/gdb/}.

\bibitem{ekberg15}
J.-E. Ekberg, ``Personal communication,'' 2015, {Trustonic}.

\bibitem{web:trustonic_t_dev}
Trustonic, ``‹t-dev developer program,''
  \url{https://www.trustonic.com/products-services/developer-program/}.

\bibitem{bookchapter:sus}
J.~Brooke, \emph{Usability evaluation in industry}.\hskip 1em plus 0.5em minus
  0.4em\relax Taylor \& Francis, London, 1996, ch. SUS: A ``quick and dirty"
  usability scale, pp. 189--194.

\bibitem{jour:BKM08}
A.~Bangor, P.~T. Kortum, and J.~T. Miller, ``An empirical evaluation of the
  system usability,'' \emph{International Journal of Human-Computer
  Interaction}, pp. 574--594, 2008,
  \url{http://dx.doi.org/10.1080\%2F10447310802205776}.

\bibitem{thesis:QEMU}
A.~Muthu, ``Emulating trust zone feature in android emulator by extending
  qemu,'' Master's thesis, KTH Royal Institute of Technology, 2013.

\bibitem{WinterWPT11}
\BIBentryALTinterwordspacing
J.~Winter, P.~Wiegele, M.~Pirker, and R.~T{\"{o}}gl, ``A flexible software
  development and emulation framework for {ARM} {TrustZone},'' in \emph{Trusted
  Systems - Third International Conference, {INTRUST} 2011, Beijing, China,
  November 27-29, 2011, Revised Selected Papers}, 2011, pp. 1--15. [Online].
  Available: \url{http://dx.doi.org/10.1007/978-3-642-32298-3_1}
\BIBentrySTDinterwordspacing

\bibitem{web:NVIDIA_TLK}
NVIDIA, ``Trusted little kernel (tlk),''
  \url{http://nv-tegra.nvidia.com/gitweb/?p=3rdparty/ote_partner/tlk.git;a=summary}.

\bibitem{web:linaro_op_tee}
Linaro, ``{OP-TEE},''
  \url{https://wiki.linaro.org/WorkingGroups/Security/OP-TEE}.

\bibitem{web:T6}
TrustKernel, ``T6,'' \url{http://trustkernel.org/}.

\bibitem{web:openVirtua}
Sierraware, ``Open virtualization's sierravisor and {SierraTEE},''
  \url{http://www.openvirtualization.org/}.

\bibitem{web:Apache-v2}
{The Apache Software Foundation}, ``Apache license, version 2.0,''
  \url{http://www.apache.org/licenses/LICENSE-2.0}.

\end{thebibliography}

% That's all folks!
\end{document}